\documentclass[twocolumn]{aastex701}

\usepackage{enumitem,graphicx,amsmath,microtype,placeins}

\newcommand{\Ha}{H\ensuremath{\alpha}}

\begin{document}

\shorttitle{UV--Optical Diagnostic for Rejuvenating Galaxies}
\shortauthors{Lazarus \& Parker}

\title{An Improved UV--Optical Diagnostic for Rejuvenating Galaxies in the Local Universe and Implications for Galaxy Evolution}

\correspondingauthor{Dylan Lazarus}
\email{lazarusd@mcmaster.ca}

\author[0009-0005-2987-7505]{Dylan Lazarus}
\affiliation{Department of Physics and Astronomy, McMaster University, 1280 Main Street West, Hamilton, ON, L8S 3L8, Canada}
\email{lazarusd@mcmaster.ca}

\author[0000-0003-4722-5744]{Laura C. Parker} 
\affiliation{Department of Physics and Astronomy, McMaster University, 1280 Main Street West, Hamilton, ON, L8S 3L8, Canada}
\email{lparker@mcmaster.ca}

\begin{abstract}

Rejuvenating galaxies are important probes of galaxy evolution, yet identifying them observationally is challenging as constraining recent star formation histories requires both photometric and spectroscopic data. We present a method for identifying rejuvenating galaxies in the local Universe using ultraviolet (UV) imaging and optical spectroscopy, building on a recent selection that identifies a system as rejuvenating if it is quenched in the near-UV (NUV; tracing $\sim\!100\,\mathrm{Myr}$ timescales) but star-forming in \Ha\ (tracing $\sim\!10\,\mathrm{Myr}$ timescales). Shortly after a star formation episode, however, the NUV is dominated by the same massive stars that power \Ha, so these indicators do not always trace distinct timescales. To address this, we derive a relation that predicts the NUV emission associated with the ionizing O-star population traced by \Ha, enabling us to isolate the NUV contribution from longer-lived stars (primarily B/A stars with $M\lesssim\!20\,M_\odot$). Subtracting the predicted O-star NUV from the dust-corrected NUV yields a more reliable rejuvenation diagnostic. Using this method, we identify $\sim\!10^{4}$ rejuvenating galaxies in a sample of Sloan Digital Sky Survey (SDSS) galaxies ($\sim\!4.5\%$). These galaxies have intermediate stellar masses and are found primarily in lower-density environments, becoming increasingly rare toward the centers of groups and clusters. Rejuvenating galaxies also exhibit systematically lower gas-phase metallicities, consistent with fueling by the accretion of metal-poor gas.

\end{abstract}


\section{Introduction} \label{sec:intro}

Galaxies are largely bimodal in their properties, commonly classified as either blue and star-forming or red and quenched \citep{Strateva2001, Baldry2004, Bell2004}. Over cosmic time, star formation rates (SFRs) decline on average \citep{Madau2014, Speagle2014}, and many galaxies transition from the star-forming main sequence (SFMS) to quiescence, passing through the so-called green valley. This sparsely populated region in color--magnitude space is associated with suppressed SFRs \citep{Schiminovich2007, Wyder2007, Schawinski2014}, and the green-valley phase itself is relatively short-lived, typically lasting tens of megayears to a gigayear depending on the quenching mechanism \citep{Salim2014, Schawinski2014}.

Galaxies can be quenched through a variety of internal and external processes that remove or heat the cold gas necessary for star formation \citep[e.g.,][]{Peng2010}. These mechanisms include feedback from stars and active galactic nuclei \citep[AGN; e.g.,][]{Croton2006, Hopkins2014}, morphological stabilization of gas against fragmentation \citep{Martig2009}, and environmental processes such as ram-pressure stripping and tidal interactions \citep{Gunn1972, Dressler1983, Poggianti2017, Trussler2020}.

Although quenching has traditionally been viewed as irreversible \citep{Sandage1961, Tinsley1972, vanDokkum2001}, recent studies suggest that some galaxies undergo rejuvenation, where star formation is reactivated following a period of quiescence. While early-type galaxies (ETGs) are typically red and passive, a minority exhibit blue optical colors and nebular emission lines \citep{Rose1985}, indicating low-level star formation activity also seen in the infrared (IR)/UV \citep{Burstein1988}. With Galaxy Evolution Explorer \citep[GALEX;][]{Martin2005}, surveys of morphologically selected local ETGs have shown that over $\sim30\%$ of these systems exhibit signatures of recent ($\sim\!10^8$--$10^9\,\mathrm{yr}$) star formation that is often invisible at optical wavelengths \citep{Kaviraj2007, Rampazzo2007, Schawinski2007}.

More recent work finds rejuvenation most commonly in lower-mass ($\lesssim\!10^{11}\,M_\odot$), disky galaxies in low- to intermediate-density environments \citep{Chauke2019, Tanaka2024}. \citet{Cleland2021} (hereafter CM21) reported similar rejuvenation fractions in field and group galaxies, consistent with models in which gas reaccretion is the primary driver rather than large-scale environment \citep{Fortune2025}. Proposed physical drivers of rejuvenation include external gas accretion via minor mergers or filaments \citep{Keres2005, Kaviraj2009, Nelson2021}, internal processes such as bar-driven inflows \citep{James2009}, and recycled gas from evolved stars \citep{Leitner2011}, potentially modulated by AGN feedback \citep{aird2019, MartinNavarro2022, woodrum2024}. Rejuvenation complicates interpretation of green-valley systems (which host both quenching and rejuvenating populations; \citealp{Fang2012, Chauke2019}) and can bias inferred stellar ages and metallicities from integrated spectra \citep{Carnall2019, Angthopo2020}. Cosmological simulations and empirical models (e.g., \textsc{IllustrisTNG}, \textsc{UniverseMachine}, \textsc{Magneticum}) predict that up to $\sim\!70\%$ of massive galaxies experience at least one rejuvenation episode by $z\!=\!0$ \citep{Nelson2018, Pillepich2018, Behroozi2019, Fortune2025}, underscoring the need for robust observational identification.

Spectral energy distribution (SED) fitting can recover star formation histories (SFHs) from multi-wavelength photometry \citep[e.g.,][]{Carnall2018, Akhshik2021, Zhang2023} and has revealed substantial populations of candidate rejuvenated systems across cosmic time: \citet{Chauke2019} found $\sim\!16\%$ of quiescent galaxies at $z\!\sim\!0.8$ experienced secondary episodes lasting $\sim\!0.7\,\mathrm{Gyr}$ and contributing $\sim\!10\%$ of their stellar mass; \citet{Tanaka2024} identified $>\!10^3$ rejuvenated galaxies at $z\!\simeq\!0$ that together contribute $\sim\!17\%$ of the local SFR density despite comprising $\lesssim\!0.1\%$ of the stellar mass. However, SED-based approaches require extensive data and are susceptible to outshining and age--dust--metallicity degeneracies \citep{Papovich2001, Maraston2010, Pforr2012, Simha2014, Leja2019}. Additionally, their coarse temporal resolution makes it difficult to detect brief star formation episodes. Simpler, scalable methods that use widely available UV and spectroscopic metrics remain highly desirable.

CM21 introduced a simplified rejuvenation selection based solely on GALEX NUV and \Ha, labeling galaxies with strong \Ha\ but weak NUV as likely rejuvenating systems. However, this approach assumes fixed timescales traced by both star formation tracers, does not correct for dust attenuation, and implicitly treats the NUV continuum as dominated by B/A-stars, even though O-stars provide most of the NUV emission during the youngest star formation episodes. In this study, we apply consistent dust corrections and explicitly separate the short- and intermediate-timescale signals traced by \Ha\ and NUV. As O-stars contribute substantially to the NUV continuum during recent star-formation episodes, we use \Ha\ as a tracer of the O-star population to predict their associated NUV emission and subtract it from the observed NUV, isolating the component produced by longer-lived stellar populations. Our approach retains the simplicity of GALEX+SDSS inputs while improving classification fidelity, enabling a comprehensive analysis of rejuvenating galaxies across environment, morphology, and stellar mass without full SED fitting and with reduced sensitivity to outshining from young stellar populations.

This paper is structured as follows. In Section~\ref{sec:data} we describe the data sets and sample selection. Section~\ref{sec:method} outlines our methodology. Results are presented in Section~\ref{sec:results} and discussed in Section~\ref{sec:discussion}. We summarize in Section~\ref{sec:summary}. We assume a flat $\Lambda$CDM cosmology with $H_0\!=\!70\,\mathrm{km\,s^{-1}\,Mpc^{-1}}$, $\Omega_{\mathrm{M}}\!=\!0.3$, and $\Omega_{\Lambda}\!=\!0.7$ throughout.

\section{Data}\label{sec:data}

We use a large, low-redshift SDSS galaxy sample to test our identification method and to study the properties of the selected samples.

\subsection{GALEX--SDSS--WISE Legacy Catalog}
Our parent sample is drawn from the GALEX–SDSS–WISE Legacy Catalog \citep[GSWLC;][]{Salim2016}, which provides a cross-match between GALEX, SDSS, and WISE photometry for SDSS spectroscopic targets. For additional datasets that do not provide SDSS object identifiers, we perform positional matching within a $1\arcsec$ radius, after testing to confirm that this radius provides reliable matches. We adopt the GSWLC--A subset to maximize areal coverage (relative to the medium-depth GSWLC--M), accepting slightly lower per-object UV signal-to-noise ratio (S/N) in exchange for a larger UV-detected sample.

To mitigate NUV non-detections in the GALEX pipeline photometry (often caused by close pairs or bright-star contamination), we incorporate forced-photometry NUV magnitudes from EMphot \citep{Osborne2023}. EMphot employs a more conservative detection threshold and excludes unreliable fits, which yields more non-detections than the pipeline despite higher photometric accuracy for detected sources. For sources detected by both pipelines, the NUV magnitudes show tight agreement with an intrinsic scatter of $\sim0.3$~mag around the one-to-one line, confirming that the two are consistent within the photometric uncertainties. Accordingly, we adopt EMphot measurements when available and retain pipeline fluxes otherwise.

We select galaxies with $z<0.1$ to ensure rest-frame NUV coverage \citep{Wyder2007} while minimizing evolutionary effects, and we impose a stellar-mass cut of $\log(M_\star/M_\odot)>9.0$ for reasonable completeness while excluding dwarf galaxies. A fully stellar-mass--complete sample within this redshift range would require $\log(M_\star/M_\odot)\gtrsim10$. We adopt a lower threshold to increase the sample size for our statistical analysis.

\subsection{WISE Mid-infrared Photometry}

To correct the NUV for dust attenuation (Section~\ref{sec:dust}), we use mid-infrared (MIR) measurements from the Wide-field Infrared Survey Explorer \citep[WISE;][]{Wright2010} W3 ($12\,\mu\mathrm{m}$) and W4 ($22\,\mu\mathrm{m}$) bands, which trace polycyclic aromatic hydrocarbon (PAH) emission and dust-obscured star formation. MIR fluxes are from the unWISE forced-photometry catalog \citep{Lang2014}, which yields higher signal-to-noise and fewer non-detections than standard WISE catalog photometry.

\subsection{SDSS Optical Spectroscopy}

Optical spectroscopic quantities, including \Ha\ and H$\beta$ line fluxes and equivalent widths, are taken directly from the Max Planck Institute for Astrophysics–Johns Hopkins University (\textsc{MPA--JHU}) value-added catalog \citep{Brinchmann2004} based on SDSS Data Release 7 (DR7) \citep{Abazajian2009}. In \textsc{MPA--JHU}, $\mathrm{EW}_{\Ha}$ is measured from the deblended \Ha\ line using a local continuum fit in the SDSS fiber spectrum.

To exclude AGN- or shock-dominated emission, we classify galaxies using the three standard Baldwin–Phillips–Terlevich (BPT) diagrams: [\ion{O}{3}]$\lambda5007$/H$\beta$ versus [\ion{N}{2}]$\lambda6584$/\Ha, [\ion{S}{2}]$(\lambda6717+\lambda6731)$/\Ha, and [\ion{O}{1}]$\lambda6300$/\Ha\ \citep{Baldwin1981}. We adopt the \citet{Kauffmann2003} empirical star-forming boundary and retain only galaxies that have $\mathrm{S/N}\!\geq\!3$ in all required emission lines and fall within the star-forming region in all three diagrams. This selection ensures that the measured \Ha\ predominantly traces star formation, allowing a reliable conversion to the expected NUV from the ionizing O-star population (Section~\ref{sec:method-main}).

SDSS spectroscopy samples only the central $3\arcsec$ of each galaxy. Therefore, we aperture-correct \Ha\ luminosities to total values using the $r$-band continuum scaling described by \citet{Brinchmann2004} and \citet{Hopkins2003}. This method assumes that the \Ha\ and $r$-band fluxes have the same fiber-to-total ratio and scales the measured (fiber) \Ha\ luminosity by the ratio of total to fiber $r$-band flux. As a robustness check against large aperture corrections, we repeat the analyses in Sections~\ref{sec:results} and \ref{sec:discussion}, restricting to galaxies with $|m_{r,\mathrm{total}} - m_{r,\mathrm{fiber}}| < 1$ mag. The results are unaffected by this cut, confirming that our conclusions are not driven by aperture effects.

The $z\!<\!0.1$, $\log(M_\star/M_\odot)\!>\!9.0$ GSWLC--A sample contains $223{,}703$ galaxies, which we refer to as the parent sample throughout this work.

\subsection{Yang Group and Cluster Catalog}\label{sec:yang}

We obtain environmental properties from the \citet{Yang2007} SDSS DR7 group catalog. We define groups as halos with $10^{13}\,M_\odot\!\le\!M_h\!<\!10^{14}\,M_\odot$ and clusters as $M_h\!\ge\!10^{14}\,M_\odot$, requiring at least three members per system. We define field galaxies as isolated centrals in the \citet{Yang2007} catalog with no detected satellites ($N=1$) and halo masses $M_h<10^{13}\,M_\odot$. For each matched halo we compute $r_{180}$ using Equation (5) of \citet{Yang2007}:
\begin{equation}
  r_{180}
  = 1.26\,h^{-1}\,\mathrm{Mpc}\,
    \left(\frac{M_h}{10^{14}\,h^{-1}M_\odot}\right)^{1/3}
    (1+z_{\mathrm{group}})^{-1}.
\end{equation}

Environment properties are available for a subset of the parent sample ($N_{\rm env}\!=\!216{,}684$). Objects not matched to the \citet{Yang2007} catalog or located outside its footprint are excluded from environment-dependent analyses. The environment-matched sample contains $153{,}659$ field galaxies ($\sim\!71\%$), $26{,}685$ group galaxies ($\sim\!12\%$), and $13{,}219$ cluster galaxies ($\sim\!6\%$), with the remaining $\sim\!11\%$ falling outside these classifications. This Yang-matched subset is used exclusively for the environmental analysis in Section~\ref{sec:results_env} and for environment-specific discussion in Section~\ref{sec:discussion}.

\section{Methodology}\label{sec:method}

\subsection{The CM21 UV--Optical Selection of Rejuvenating Galaxies}\label{sec:orig_method}

We first summarize the UV--optical selection of CM21, which serves as the starting point for our method. This approach compares two recent star formation diagnostics that differ in temporal sensitivity because they trace different, though overlapping, stellar populations:
\begin{enumerate}[itemsep=1ex, leftmargin=*]
    \item \textbf{NUV--based specific SFR ($\mathrm{sSFR}_{\mathrm{NUV}}$):}
    An estimate of the recent star formation rate averaged over the past $\sim\!100\,\mathrm{Myr}$, derived from the GALEX NUV continuum. CM21 converts the NUV luminosity to a star formation rate using the \citet{Kennicutt1998} calibration
    \begin{equation}
        \mathrm{SFR}_{\mathrm{NUV}} = 1.4\times10^{-28}\,L_\nu \quad [M_\odot\,\mathrm{yr^{-1}}],
    \end{equation}
    with $L_\nu$ in $\mathrm{erg\,s^{-1}\,Hz^{-1}}$, assuming a continuous star formation history and a \citet{Salpeter1955} initial mass function (IMF). The corresponding specific star formation rate is then defined as $\mathrm{sSFR}_{\mathrm{NUV}}\!\equiv\!\mathrm{SFR}_{\mathrm{NUV}}/M_\star$. For typical star-forming galaxies, the NUV emission is dominated by intermediate-mass B/A-type stars with lifetimes of $\sim\!10^8\,\mathrm{yr}$.
    
    \item \textbf{\Ha\ equivalent width ($\mathrm{EW}_{\Ha}$):}
    We adopt the rest-frame \Ha\ equivalent widths provided by the \textsc{MPA-JHU} catalog (Section~\ref{sec:data}), defined as the \Ha\ line flux divided by the underlying continuum flux density at \Ha. In non-AGN systems, $\mathrm{EW}_{\Ha}$ traces $\sim\!10\,\mathrm{Myr}$ timescales as it is dominated by short-lived massive O- and B-type stars \citep[e.g.,][]{Kennicutt2012}. As a line-to-continuum ratio, $\mathrm{EW}_{\Ha}$ serves as a proxy for $\mathrm{sSFR}_{\Ha}$ and is less sensitive than line flux to dust and aperture effects \citep[e.g.,][]{Kennicutt2012}.
\end{enumerate}

\begin{figure}
    \centering
    \includegraphics[width=\columnwidth]{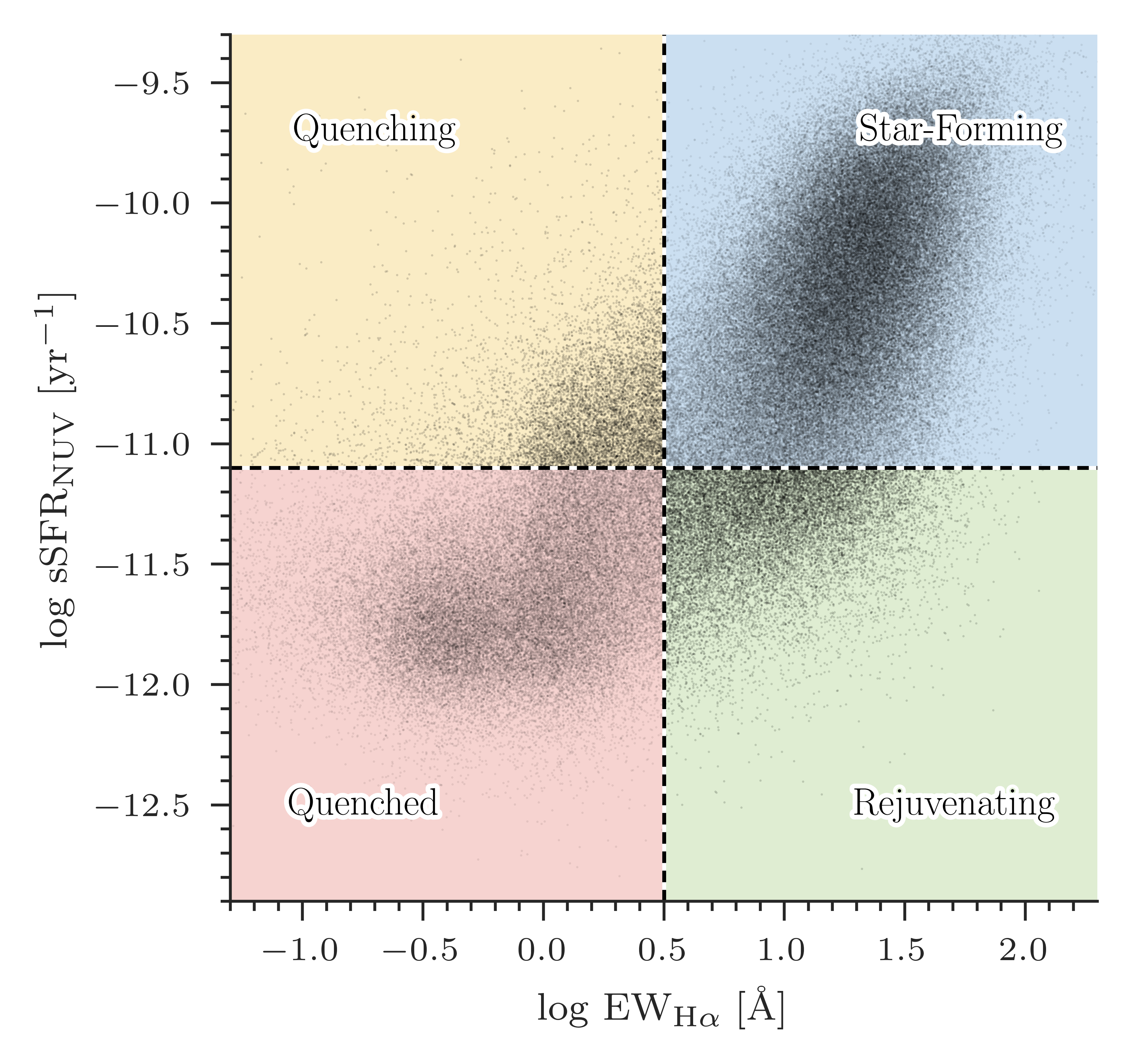}
    \caption{Distribution of the GSWLC parent sample in $\mathrm{EW}_{\Ha}$ vs. $\mathrm{sSFR}_{\mathrm{NUV}}$ space, illustrating four subpopulations---star-forming (blue), rapidly quenching (yellow), quenched (red), and rejuvenating (green)---as defined by CM21. The boundaries separating active star formation from quiescence are set at $\log_{10}\!\big(\mathrm{EW}_{\Ha}/\text{\AA}\big)=0.5$ and $\log_{10}\!\big(\mathrm{sSFR}_{\mathrm{NUV}}/\mathrm{yr}^{-1}\big)=-11.1$, corresponding to minima in the respective relations with stellar mass.}
    \label{fig:ew_versus_ssfr}
\end{figure}

Comparing $\mathrm{sSFR}_{\mathrm{NUV}}$ and $\mathrm{EW}_{\Ha}$ (or, equivalently, $\mathrm{sSFR}_{\Ha}$), CM21 classify galaxies into four categories:
\begin{enumerate}[label=(\roman*), itemsep=1ex, leftmargin=*]
    \item \textit{Star-forming}: high $\mathrm{sSFR}_{\mathrm{NUV}}$ and high $\mathrm{EW}_{\Ha}$, indicating sustained activity over the past $\sim\!100\,\mathrm{Myr}$ and at present;
    \item \textit{Quenched}: low in both indicators, indicating little to no recent star formation;
    \item \textit{Quenching}: high $\mathrm{sSFR}_{\mathrm{NUV}}$ but low $\mathrm{EW}_{\Ha}$, consistent with significant activity tens of megayears ago but little ongoing formation;
    \item \textit{Rejuvenating}: low $\mathrm{sSFR}_{\mathrm{NUV}}$ but high $\mathrm{EW}_{\Ha}$, implying a formerly quiescent system with newly reignited star formation.
\end{enumerate}

In the $\mathrm{sSFR}_{\mathrm{NUV}}$--$\mathrm{EW}_{\Ha}$ plane (Figure~\ref{fig:ew_versus_ssfr}), rejuvenating systems occupy the bottom-right region with $\log_{10}\!\big(\mathrm{EW}_{\Ha}/\text{\AA}\big)\!\ge\!0.5$ but $\log_{10}\!\big(\mathrm{sSFR}_{\mathrm{NUV}}/\mathrm{yr}^{-1}\big)\!<\!-11.1$, while rapidly quenching systems occupy the top-left with $\log_{10}\!\big(\mathrm{EW}_{\Ha}/\text{\AA}\big)\!<\!0.5$ but $\log_{10}\!\big(\mathrm{sSFR}_{\mathrm{NUV}}/\mathrm{yr}^{-1}\big)\!\ge\!-11.1$.

Our follow-up analysis highlights two limitations of the CM21 selection. \textit{(i) Dust:} the UV is attenuated more strongly than optical lines \citep{Calzetti2000,Battisti2016}, so dusty star-forming galaxies can show depressed NUV relative to \Ha\ and be misclassified as rejuvenating \citep[e.g.,][]{Hao2011}. \textit{(ii) Overlapping timescales:} UV and \Ha\ are not independent tracers; a given episode of star formation produces both, and shortly after an episode, the NUV continuum is dominated by the same massive stars that power \Ha. Robust identification of rapid and recent temporal changes in SFR, therefore, requires disentangling these indicators.

In Sections~\ref{sec:dust} and \ref{sec:method-main}, we correct for dust and disentangle timescales by predicting the NUV emission associated with the O-star population traced by \Ha\ and subtracting it from the attenuation-corrected NUV.

\subsection{Dust Attenuation Corrections}\label{sec:dust}

\begin{figure}
    \centering
    \includegraphics[width=\columnwidth]{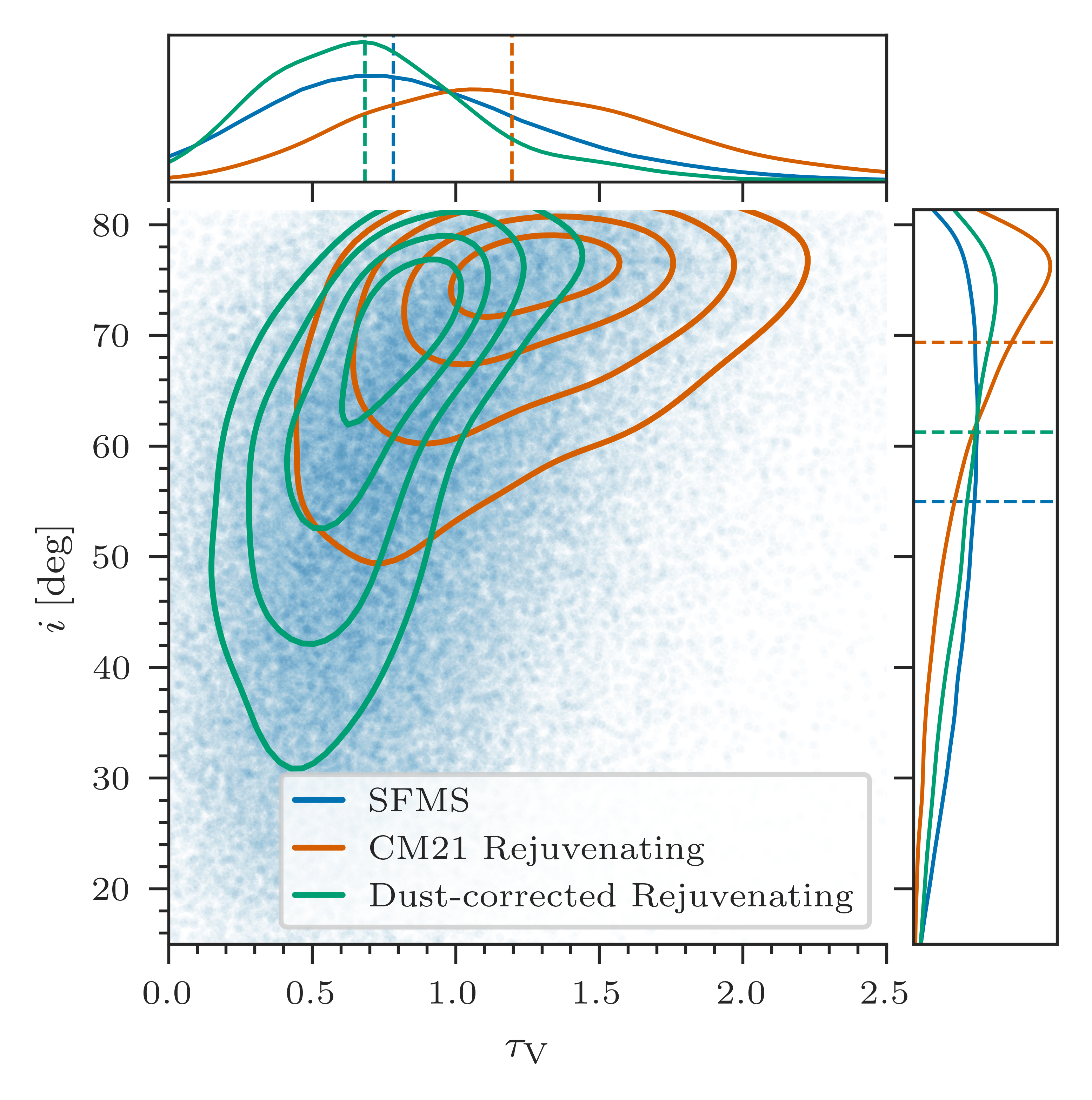}
    \caption{Inclination $i$ from axis ratio following \citet{Blanton2011} vs. $V$-band optical depth $\tau_V$ from \textsc{MPA--JHU} for the CM21 rejuvenating sample (red) and our dust-corrected rejuvenating sample (green). The faint blue points show the distribution of star-forming GSWLC--A galaxies. Kernel density contours are shown for both rejuvenating samples. Marginal kernel density estimates (KDEs) along the axes summarize $i$ and $\tau_V$. Dashed lines mark medians.}
    \label{fig:inc_vs_tau}
\end{figure}

Figure~\ref{fig:inc_vs_tau} compares photometric inclination $i$ and $V$-band optical depth $\tau_V$ for both the CM21 selection and for our sample. We obtain $i$ from the NSA single-S\'ersic axis ratio $q=b/a$ (\texttt{SERSIC\_BA}) provided by the NASA--Sloan Atlas \citep{Blanton2011}, converting $q$ to inclination with the standard oblate-disk relation and an intrinsic thickness $q_0$ \citep[e.g.,][]{Padilla2008}. The $\tau_V$ values are taken from the \textsc{MPA--JHU} DR7 emission-line modeling \citep{Brinchmann2004}, which follows the attenuation model of \citet{Charlot2001}. In Figure~\ref{fig:inc_vs_tau}, we see that the CM21 rejuvenating candidates are concentrated toward high $i$ and high $\tau_V$.

This trend arises because edge-on sightlines increase the path length through the dusty disk midplane, suppressing the NUV continuum more strongly than \Ha\ and lowering $\mathrm{sSFR}_{\mathrm{NUV}}$, which moves dusty disks into the ``low-NUV/high-\Ha'' quadrant and incorrectly identifies galaxies as rejuvenating. We correct the photometry using an infrared-excess (IRX)--based method and correct \Ha\ using the Balmer decrement, as described below.

\medskip
\noindent\textbf{NUV continuum:}
We estimate the NUV attenuation, $A_{\mathrm{NUV}}$, using the IRX method, which relates the ratio of total infrared (TIR) to ultraviolet emission to dust obscuration. We compute $L_{\mathrm{TIR}}$ from WISE mid-infrared bands using the empirical calibrations of \citet{Cluver2017}:
\begin{align}
  \log_{10}\!\left(\frac{L_{\mathrm{TIR}}}{L_\odot}\right) &= 0.889\,\log_{10}\!\left(\frac{L_{\mathrm{PAH}}}{L_\odot}\right) + 2.21, \\
  \log_{10}\!\left(\frac{L_{\mathrm{TIR}}}{L_\odot}\right) &= 0.915\,\log_{10}\!\left(\frac{L_{\mathrm{dust}}}{L_\odot}\right) + 1.96,
\end{align}
where $L_{\mathrm{PAH}}$ is the stellar--continuum--corrected $12\,\mu\mathrm{m}$ luminosity (W3) and $L_{\mathrm{dust}}$ is the $22\,\mu\mathrm{m}$ luminosity (W4). To remove the stellar continuum from W3/W4, we subtract a scaled W1 measurement before computing $L_{\mathrm{PAH}}$ following \citet{Cluver2017}. When both are available, we use the W4-based calibration; when only W3 is available, we use the W3-based calibration. Defining $\mathrm{IRX}\equiv\log_{10}(L_{\mathrm{TIR}}/L_{\mathrm{NUV}})$ from the observed $L_{\mathrm{NUV}}$, we adopt \citet{Hao2011}:
\begin{equation}
A_{\mathrm{NUV}} = 2.5\,\log_{10}\!\bigl(1 + 0.27\times10^{\mathrm{IRX}}\bigr),
\label{eq:nuvatt}
\end{equation}
and correct NUV luminosities accordingly for use in $\mathrm{sSFR}_{\mathrm{NUV}}$. We do not consider possible mid-infrared contamination from AGN, as such systems are already removed from our sample following the BPT-based cuts described in Section~\ref{sec:data}.\\

\medskip
\noindent\textbf{\Ha\ line:}
We correct \Ha\ for dust using the Balmer decrement, the observed H$\alpha$/H$\beta$ ratio measured from SDSS DR7 emission lines \citep{Brinchmann2004}. We use stellar-absorption--corrected Balmer fluxes from the \textsc{MPA--JHU} catalog. We assume an intrinsic Balmer ratio of 2.86, corresponding to Case~B ($f_{\mathrm{esc}}=0$) with $T_e=10^4\,\mathrm{K}$ and $n_e=10^2\,\mathrm{cm}^{-3}$ \citep{Osterbrock2006}, and compute the attenuation with
\begin{equation}
F_{\mathrm{H}\alpha,\mathrm{corr}} = F_{\mathrm{H}\alpha}\,\left[\frac{(F_{\mathrm{H}\alpha}/F_{\mathrm{H}\beta})}{2.86}\right]^{\alpha},
\label{eq:balmer}
\end{equation}
where
\begin{equation}
\alpha = \frac{k_{\mathrm{H}\alpha}}{k_{\mathrm{H}\beta}-k_{\mathrm{H}\alpha}} = 2.36,
\label{eq:diff_ext}
\end{equation}
with $k_\lambda$ from \citet{Cardelli1989}, updated by \citet{ODonnell1994} (Milky Way, $R_V=3.1$). Fiber fluxes are scaled to total fluxes as described in Section~\ref{sec:data}.

\begin{figure}
    \centering
    \includegraphics[width=0.97\columnwidth]{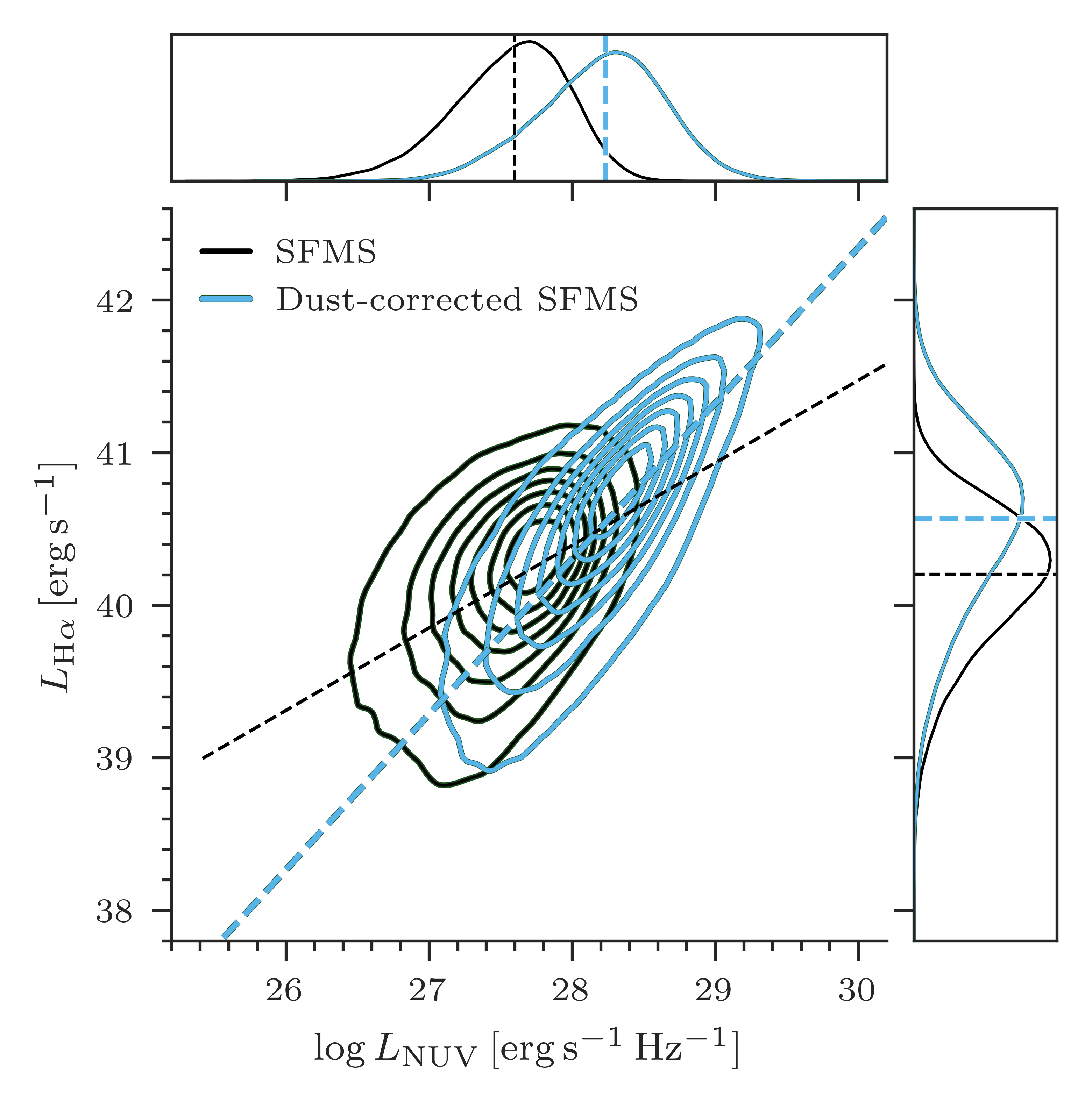}
    \caption{Kernel density distributions comparing \Ha\ and NUV luminosities before (thin black) and after (thick blue) dust corrections for the \textsc{MPA--JHU} star-forming sample \citep{Brinchmann2004}. Best-fit linear trends are overplotted and illustrate improved correlation after correction. Marginal KDEs display one-dimensional distributions. Dashed lines indicate medians.}
    \label{fig:corr_vs_uncorr}
\end{figure}

After applying these dust corrections, the rejuvenating galaxy sample no longer peaks at high inclination or high $\tau_V$. The concentration of high-$i$ and high-$\tau_V$ in the original CM21 rejuvenating sample indicates that many of those systems were in fact dusty star-forming galaxies rather than truly rejuvenating. After correction, the $i$–$\tau_V$ distribution closely matches that of the general star-forming population (Figure~\ref{fig:inc_vs_tau}), indicating that the corrected sample captures genuinely rejuvenating systems, not dusty star-forming galaxies misidentified due to viewing angle or attenuation effects.

The resulting $A_{\mathrm{NUV}}$ values for our parent sample (see Section \ref{sec:data}) span nearly dust-free systems to heavily obscured starbursts, with $\sim\!90\%$ of galaxies having $A_{\mathrm{NUV}}\!\le\!3\,\mathrm{mag}$. As expected for typical SFMS galaxies, dust-corrected \Ha\ and NUV luminosities show tighter correspondence and reduced scatter (Figure~\ref{fig:corr_vs_uncorr}).

While dust corrections mitigate much of the bias introduced by attenuation and inclination, they do not resolve the fundamental overlap in timescales traced by \Ha\ and NUV emission. To address this, we use stellar population synthesis models to estimate and subtract the O-star contribution to the NUV continuum, thereby disentangling timescales as detailed below in Section~\ref{sec:method-main}.

\subsection{Main Calibration and Motivation}\label{sec:method-main}

In a typical star-forming galaxy with a roughly constant SFR over extended periods, nebular \Ha\ traces the instantaneous ionizing-photon production rate from O stars on $\lesssim\!5$--$10\,\mathrm{Myr}$ timescales, whereas the NUV continuum is dominated by B stars over $\sim\!10$--$100\,\mathrm{Myr}$. Immediately following a rejuvenation event, \Ha\ emission rises sharply while the B-star population has not yet built up, producing a short ``pre-equilibrium'' phase in which \Ha\ appears elevated relative to the NUV. Quantifying this phase requires disentangling the NUV contribution of ionizing stars from that of longer-lived B stars.

The ionizing-star contribution to the NUV cannot be separated observationally from broadband data, so it must be estimated indirectly. To do so, we use stellar population synthesis models to calibrate the relation between $L_{\mathrm{H}\alpha}$ and the associated O-star NUV continuum.

Model $L_{\mathrm{H}\alpha}$ and NUV continua are generated using Flexible Stellar Population Synthesis \citep[\textsc{FSPS};][]{Conroy2009,Conroy2010} with nebular continuum and lines enabled \citep{Byler2017}, adopting MESA Isochrones and Stellar Tracks \citep[MIST;][]{Choi2016,Dotter2016} and disabling internal dust so that attenuation is applied uniformly in the observational corrections (Section~\ref{sec:dust}). The \textsc{FSPS} nebular module is run under Case~B recombination with default electron temperature and density. We explore recent SFHs including constant, delayed-$\tau$, exponentially declining ($\tau = 10$--$1{,}000\,\mathrm{Myr}$), and burst models across $-0.5<\log(Z/Z_\odot)<0.5$ and $-3\le\log U\le-2$. For each model, we form the NUV $L_\nu$ by summing the stellar and nebular continua before band integration.

Two NUV components are then derived:
(i) the full-IMF NUV, $L_{\mathrm{NUV}}$ (stellar$+$nebular continuum from all initial stellar masses), which is the observed quantity used in our classifier; and
(ii) an O-star proxy NUV, $L_{\mathrm{NUV,ion}}$, obtained by restricting to stellar masses above $20\,M_\odot$ and adding the NUV nebular continuum. At the high-mass end, the Kroupa and Salpeter IMFs have nearly identical slopes ($\alpha\simeq2.3$–$2.35$ for $M \gtrsim 1\,M_\odot$), so our O-star--dominated \Ha–NUV conversion is insensitive to this choice \citep{Salpeter1955, Kroupa2001}.

Finally, $L_{\mathrm{NUV,ion}}$ and $L_{\mathrm{H}\alpha}$ are evaluated across the grid of star formation histories, metallicities, and ionization parameters described above, and their ratio is examined:
\begin{equation}
\label{eq:ratio}
\log_{10}\!\left(\frac{L_{\mathrm{NUV,ion}}}{L_{\mathrm{H}\alpha}}\right),
\end{equation}
where $L_{\mathrm{H}\alpha}$ is in $\mathrm{erg\,s^{-1}}$ and $L_{\mathrm{NUV,ion}}$ is a luminosity in $\mathrm{erg\,s^{-1}\,Hz^{-1}}$ corresponding to the GALEX NUV band. The ratio, therefore, has units of $\mathrm{Hz}^{-1}$. Throughout, the subscript ``ion'' denotes NUV continuum associated with the ionizing (O-star) population and does not imply that NUV photons are themselves ionizing.

\begin{figure}
    \centering
    \includegraphics[width=\columnwidth]{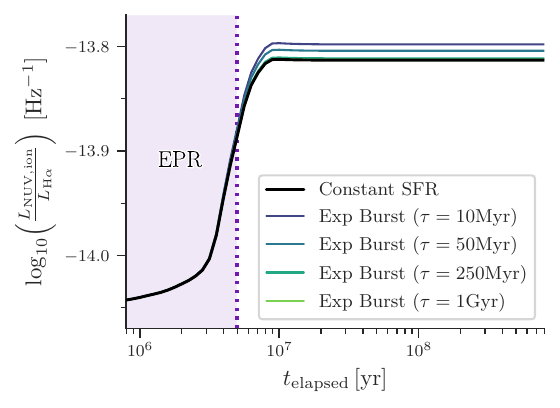}
    \caption{\textsc{FSPS} \citep{Conroy2009,Conroy2010} tracks of $\log_{10}(L_{\mathrm{NUV,ion}}/L_{\mathrm{H}\alpha})$ using the O-star NUV component and nebular H$\alpha$. The black curve shows a constant-SFR model; the colored dotted curves show exponentially declining SFHs with varying $e$-folding times. All curves converge to the equilibrium plateau of $\simeq-13.8$ after $\sim\!3$--$6\,\mathrm{Myr}$. Shading marks the EPR (pre-equilibrium) regime.}
    \label{fig:nuv_frac_tburst}
\end{figure}

Figure~\ref{fig:nuv_frac_tburst} shows this ratio for a representative set of SFHs. In all cases, the ratio rises rapidly during the first few Myr of star formation and then converges to a stable plateau of $\log_{10}(L_{\mathrm{NUV,ion}}/L_{\mathrm{H}\alpha}) \simeq -13.8$ with an intrinsic scatter of $\sim\!0.1$\,dex once the B-star continuum becomes established. Within the FSPS grid described above, the equilibrium plateau is nearly insensitive to metallicity and ionization parameter: across $-0.5 < \log(Z/Z_\odot) < 0.5$ and $-3 \le \log U \le -2$, $\log{10}(L_{\mathrm{NUV,ion}}/L_{\mathrm{H}\alpha})$ varies by at most $\sim0.1$ dex, comparable to the intrinsic scatter from varying SFHs. We therefore adopt a single equilibrium conversion factor (Equation~\ref{eq:eqconversion}).

Based on this behavior, we separate the evolution into two regimes: (i) an \textbf{equilibrium phase}, where the system has reached the plateau and is well described by the constant conversion
\begin{equation}
\label{eq:eqconversion}
L_{\mathrm{NUV,ion}}^{\mathrm{eq}} \;\approx\; 10^{-13.8}\,\mathrm{Hz}^{-1}\times L_{\mathrm{H}\alpha},
\end{equation}
and (ii) an \textbf{early-phase rejuvenation (EPR)} regime, representing the brief pre-equilibrium period immediately after rejuvenation when \Ha\ is elevated but the NUV continuum has not yet built up. A population of rejuvenating galaxies would contain both early-phase and equilibrium systems. In this work, we do not separate these two populations, but study all rejuvenating galaxies together.

We adopt the equilibrium conversion of Equation~\ref{eq:eqconversion} to remove the O-star--associated NUV and to isolate the non-ionizing NUV component tracing $10$--$100\,\mathrm{Myr}$ populations (Section~\ref{sec:method-proc}). For each galaxy we compute
\begin{equation}
    L_{\mathrm{NUV,ion}}^{\mathrm{eq}} = 10^{-13.8}\,\mathrm{Hz}^{-1} \times L_{\mathrm{H}\alpha, \mathrm{corr}}
\end{equation}
from the dust-corrected H$\alpha$ luminosity and define
\begin{equation}
    L_{\mathrm{NUV,non}}^{\mathrm{eq}} \equiv L_{\mathrm{NUV,corr}} - L_{\mathrm{NUV,ion}}^{\mathrm{eq}}.
\end{equation}
Because the equilibrium mapping slightly overpredicts the O-star NUV in systems younger than $\sim 3\,\mathrm{Myr}$ (before the B-star continuum has accumulated), objects with
$L_{\mathrm{NUV,non}}^{\mathrm{eq}} \le 0$ are in the brief pre-equilibrium phase. We flag the youngest rejuvenating systems with $\mathrm{EPR}=1$ if $L_{\mathrm{NUV,non}}^{\mathrm{eq}} \le 0$, and $\mathrm{EPR}=0$ otherwise. 

The EPR regime is a high-purity subset within the rejuvenating class: galaxies with $L_{\mathrm{NUV,non}}^{\mathrm{eq}} \le 0$ (i.e., excess H$\alpha$ relative to the equilibrium-predicted $L_{\mathrm{NUV,ion}}$) correspond to the youngest, O-star--dominated rejuvenating systems identified by our classifier.

\subsection{Rejuvenating Galaxy Classification Pipeline}\label{sec:method-proc}

Here, we describe the steps for our classification pipeline, which returns two labels: (i) a primary SFH label, and (ii) a secondary flag for EPRs.

\begin{enumerate}
\item \textbf{Dust and aperture correct.}
Compute dust-corrected luminosities, $L_{\mathrm{NUV}}^{\mathrm{corr}}$ and $L_{\mathrm{H}\alpha}^{\mathrm{corr}}$. For NUV, use an IRX-based attenuation; for \Ha, use the Balmer decrement. Convert fiber \Ha\ to total using the $r$-band total/fiber ratio (details in Section~\ref{sec:data}).

\item \textbf{Remove the O-star NUV.}
\begin{equation}
\label{eq:eqpredict}
L_{\mathrm{NUV,ion}}^{\mathrm{eq}} \;=\; 10^{-13.8}\,\mathrm{Hz}^{-1}\!\times L_{\mathrm{H}\alpha}^{\mathrm{corr}},
\end{equation}
and define the non-O-star NUV component
\begin{equation}
\label{eq:resid}
L_{\mathrm{NUV,non}}^{\mathrm{eq}} \;\equiv\; L_{\mathrm{NUV}}^{\mathrm{corr}} \;-\; L_{\mathrm{NUV,ion}}^{\mathrm{eq}} .
\end{equation}

\item \textbf{Compute $\mathrm{sSFR}_{\mathrm{NUV,non}}$.}
Using the standard NUV calibration \citep{Kennicutt1998} for monochromatic $L_\nu$ (Salpeter IMF),
\begin{equation}
\label{eq:sfrnu}
\mathrm{SFR}_{\mathrm{NUV,non}} \;=\; 1.4\times10^{-28}\, L_{\mathrm{NUV,non}}^{\mathrm{eq}}\quad [M_\odot\,\mathrm{yr}^{-1}],
\end{equation}
\begin{equation}
\label{eq:ssfrnu}
\mathrm{sSFR}_{\mathrm{NUV,non}} \;\equiv\; \frac{\mathrm{SFR}_{\mathrm{NUV,non}}}{M_\star}\quad [\mathrm{yr}^{-1}].
\end{equation}

\item \textbf{Primary classification label.}

We identify the minima separating the two peaks in the bimodal distributions of $\mathrm{EW}_{\Ha}$ and $\mathrm{sSFR}_{\mathrm{NUV,non}}$ (following CM21) on our dust-corrected, time-disentangled sample. These minima occur at $\log_{10}(\mathrm{EW}_{\Ha}/\text{\AA}) = 0.5$ and $\log_{10}(\mathrm{sSFR}_{\mathrm{NUV,non}}/\mathrm{yr}^{-1}) = -10.5$. We thus classify galaxies as star-forming (SFG), quenched (QG), rapidly quenching (RQG), and rejuvenating (RG), summarized below:
\vspace{-0.8em}
\begin{center}
\small
\setlength{\tabcolsep}{6pt}
\renewcommand{\arraystretch}{1.05}
\begin{tabular}{@{}lcc@{}}
\hline
\textbf{Class} & $\log_{10}\mathrm{EW}_{\Ha}\,[\text{\AA}]$ & $\log_{10}\mathrm{sSFR}_{\mathrm{NUV,non}}\,[\mathrm{yr}^{-1}]$ \\
\hline
SFG & $\ge 0.5$ & $\ge -10.5$ \\
QG  & $< 0.5$  & $< -10.5$  \\
RQG & $< 0.5$  & $\ge -10.5$ \\
RG & $\ge 0.5$ & $< -10.5$ \\
\hline
\end{tabular}
\end{center}

\item \textbf{Secondary EPR selection.}
The EPR flag is a binary subset label within the rejuvenating class and is meant to pick out pre-equilibrium systems (see Section~\ref{sec:method-main}). We set $\mathrm{EPR}=1$ for galaxies with $L_{\mathrm{NUV,non}}^{\mathrm{eq}} \le 0$. In total, $24.8^{+0.5}_{-0.5}\,\%$ of RGs in the parent sample are flagged as EPRs. A full machine-readable table of the rejuvenating sample, including EPR flags, is provided in the Appendix.

\end{enumerate}

\section{Results}\label{sec:results}

\subsection{Population Demographics in the $\mathrm{EW}_{\Ha}$--$M_\star$ Plane}

\begin{figure}
  \centering
  \includegraphics[width=\columnwidth]{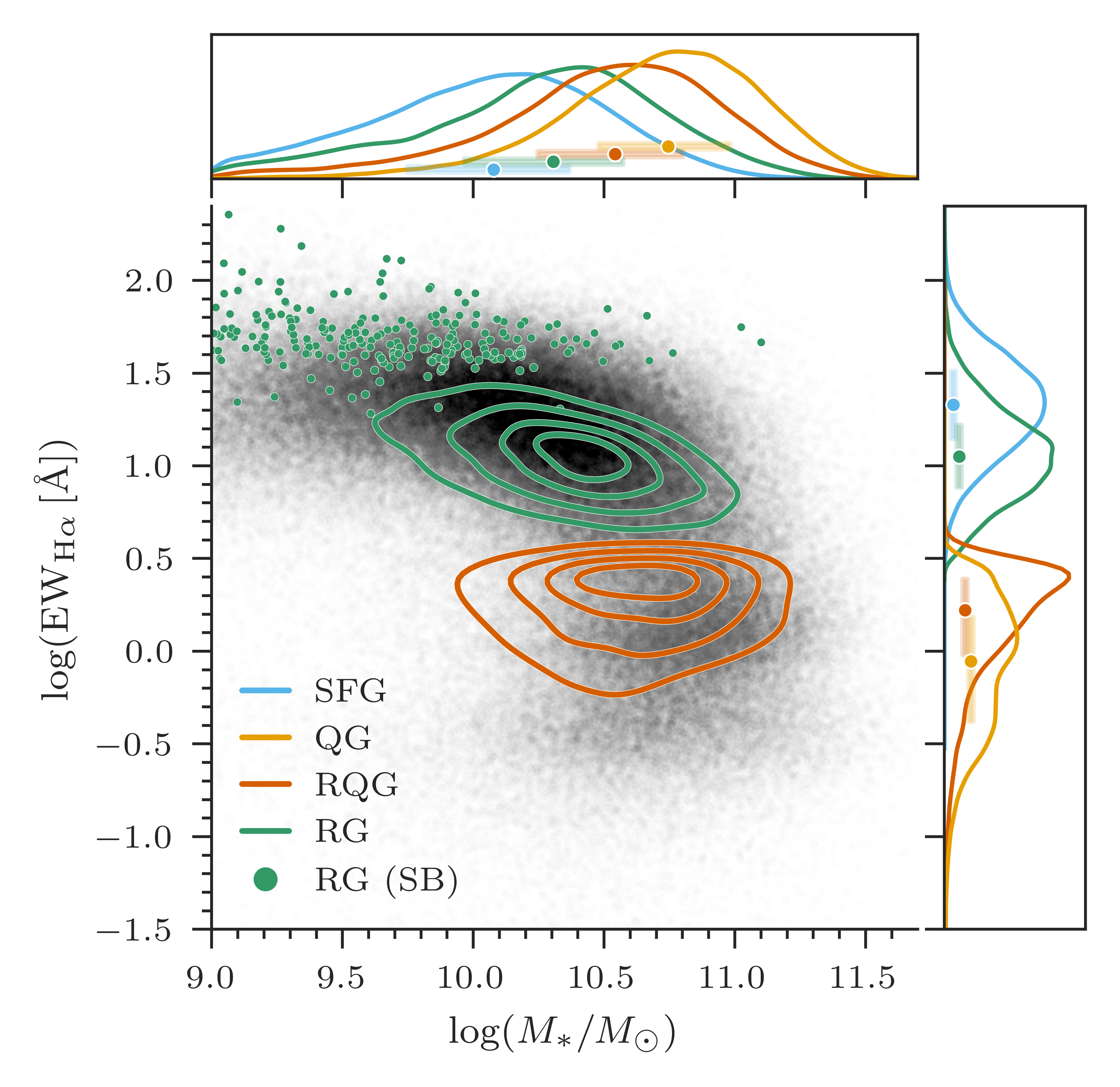}
  \caption{Distribution of galaxy subpopulations in $\mathrm{EW}_{\Ha}$ vs. stellar mass. Kernel-density contours for rejuvenating (green) and quenched (orange) systems are shown over the full parent sample (black). Green points mark RGs classified as starbursts in \textsc{MPA--JHU}. Marginal KDEs along the axes display the one-dimensional distributions (mass on the $x$-axis; $\mathrm{EW}_{\Ha}$ on the $y$-axis). Points and shaded bars denote medians and 16th--84th percentiles.}
  \label{fig:ew_mass}
\end{figure}

We first compare the distributions of the four subpopulations (SFG, QG, RQG, and RG) in the $\mathrm{EW}_{\Ha}$--$M_\star$ plane for the parent sample (Figure~\ref{fig:ew_mass}). For all results, we include all RGs (both EPR and non-EPR) unless otherwise specified. The marginal $\mathrm{EW}_{\Ha}$ distributions at fixed $M_\star$ show that RGs tend to have lower $\mathrm{EW}_{\Ha}$ than SFGs, with a median offset of $\sim\!0.25\,\mathrm{dex}$, indicating that RGs lie slightly below the SFMS. RGs are not concentrated in a narrow ``transition zone'' between the star-forming and quenched sequences; rather, their distribution is only mildly skewed below the SFMS. At the opposite extreme, we identify $\simeq 200$ RGs labeled as starbursts in the \textsc{MPA--JHU} catalog, which the dust-corrected CM21 selection misses.

\subsection{NUV--$r$ Color Evolution}\label{sec:results_color}

To test whether rejuvenating galaxies occupy the green valley, and how their colors change from before to after renewed star formation, we compare dust-corrected pre-rejuvenation colors to present colors.

We define the green valley by dividing the parent sample, excluding AGN (Section~\ref{sec:data}), into 10 stellar-mass bins spanning the full mass range. In each bin, we fit a two-component Gaussian mixture \citep[implemented via \texttt{sklearn.mixture.GaussianMixture};][]{pedregosa_2011} to the NUV--$r$ distribution and identify the minimum between the blue and red peaks. Bins without clear bimodality at low mass ($M_\star\!\lesssim\!10^{9.5}\,M_\odot$) are discarded. A linear fit to the minima defines the ridgeline of the green valley, and we adopt a $\pm0.5\,\mathrm{mag}$ band around this line as our green valley definition. Present-day colors use the IRX-corrected GALEX NUV from Equation~\ref{eq:nuvatt} with the SDSS $r$-band. As $r$-band evolution over $\lesssim\!100\,\mathrm{Myr}$ is negligible for our purposes compared to NUV variations, we assume a constant $r$ and compute pre-rejuvenation colors using the non-ionizing NUV component $L_{\mathrm{NUV,non}}^{\mathrm{eq}}$ (Equation~\ref{eq:resid}), obtained by subtracting the O-star NUV contribution from the IRX-corrected NUV (Equation~\ref{eq:eqconversion}).

\begin{figure}
  \centering
  \includegraphics[width=0.97\columnwidth]{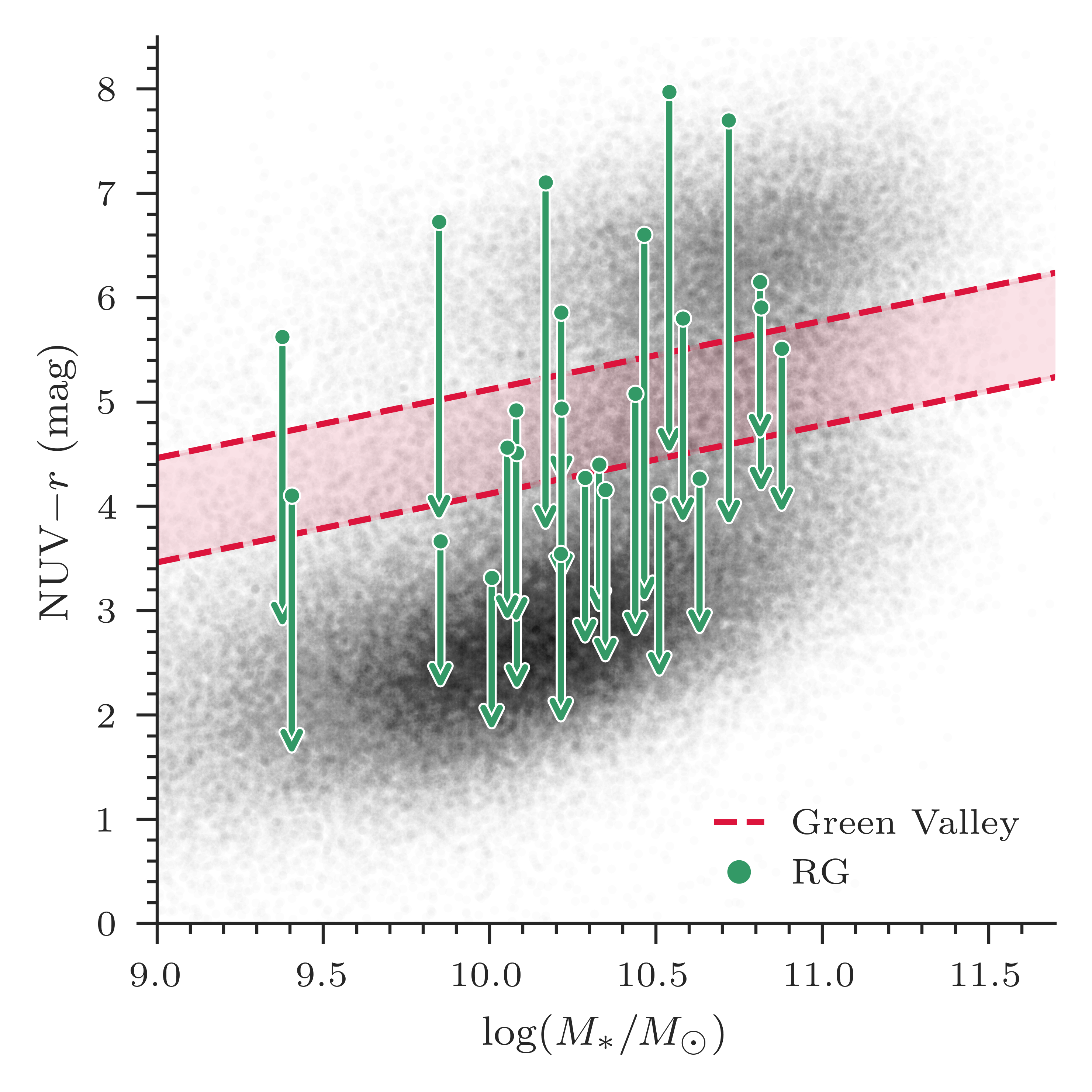}
  \caption{NUV--$r$ color evolution of rejuvenating galaxies. Points mark inferred pre-rejuvenation colors from the residual $L_{\mathrm{NUV,non}}^{\mathrm{eq}}$ (Equation~\ref{eq:resid}). Arrows point to present colors that use IRX-corrected GALEX NUV and SDSS $r$. The background shows the distribution of the parent sample. The shaded red band with dashed outline marks the green valley.}
  \label{fig:color_evolution}
\end{figure}

Figure~\ref{fig:color_evolution} illustrates a random subset of 25 non-EPR rejuvenating galaxies. Points mark inferred pre-rejuvenation colors, and arrows point to present, dust-corrected colors. The mean color change for systems with significant pre-rejuvenation NUV emission is $\sim\!0.8\,\mathrm{mag}$. Only $14.6\%$ of rejuvenating galaxies lie in the green valley at present, while most occupy the blue cloud, challenging the notion that transitioning systems are most commonly found in the green valley (see Section \ref{sec:properties}).

\subsection{Environmental Dependence}\label{sec:results_env}
\begin{figure*}
  \centering
  \includegraphics[width=\linewidth]{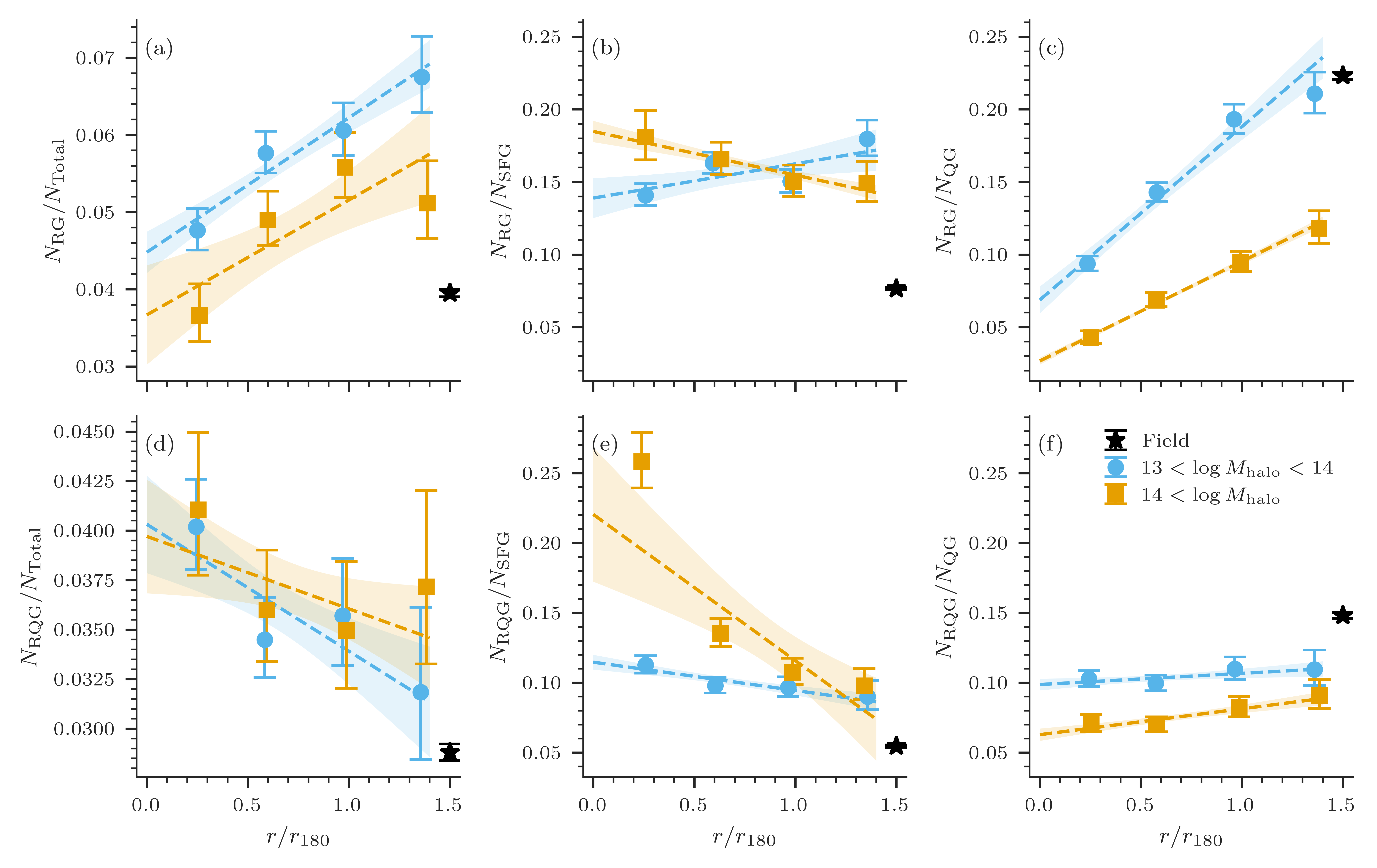}
  \caption{Fraction of rejuvenating (top row) and rapidly quenching (bottom row) galaxies relative to (a) the full sample, (b) the star-forming subset, and (c) the quenched subset, as a function of group- or cluster-centric radius $r/r_{180}$ for the Yang-matched subsample. Points show bin means with binomial errors. Lines show weighted fits. Groups are shown in blue, clusters in orange, and the field in black, plotted at $r/r_{180}=1.5$ for display purposes.}
  \label{fig:env_fraction}
\end{figure*}

We examine environmental trends using only the Yang-matched subsample, defining field, group, and cluster environments as described in Section~\ref{sec:yang}. Analogous to CM21 Figure~8, we plot the fractions of RGs and RQGs relative to (a) the full sample, (b) the SFG subset, and (c) the QG subset as a function of group- or cluster-centric radius $r/r_{180}$ (Figure~\ref{fig:env_fraction}). For groups and clusters, we bin in four equal-width intervals of $r/r_{180}$ and measure the ratio in each bin. We fit radial trends with weighted least squares, using binomial uncertainties on the ratios \citep{Cameron2011} as weights.

RG satellites are most common in the outskirts of groups, whereas RQG systems peak toward group and cluster centers. RGs occupy projected cluster-centric radii similar to those of SFGs but are comparatively underrepresented in the field. RQGs are especially prevalent in groups and clusters, consistent with environmentally driven rapid quenching processes.

\section{Discussion}\label{sec:discussion}

We relate our empirical trends to theoretical expectations from hydrodynamical simulations and to previous observational work, examining which physical mechanisms can account for the mass, color, and environmental dependencies reported here.

\subsection{Properties of Rejuvenating Galaxies and Their Environments}\label{sec:properties}

Figure~\ref{fig:ew_mass} shows that about $70\%$ of rejuvenating galaxies (RGs) have $M_\star<10^{10.5}\,M_\odot$. The stellar-mass dependence likely reflects, on average, halo-mass--dependent gas cooling. In massive halos, deep potential wells, high velocity dispersions, and hot atmospheres inhibit the cooling and resupply of circumgalactic gas \citep{Birnboim2003}. At lower $M_\star$, residual \ion{H}{1} reservoirs of $\sim\!10^{8}$--$10^{9}\,M_\odot$ (routinely detected in optically red spirals by the Arecibo Legacy Fast ALFA survey, ALFALFA) can be stirred or reaccreted, fueling brief rejuvenation episodes \citep{Giovanelli2005}.

RGs have a stellar-mass distribution that is intermediate between that of SFGs and QGs, and it is shifted to lower $M_\star$ than the QGs. This is consistent with rejuvenation being less likely at high $M_\star$, mirroring the declining SFG fraction in massive galaxies. In addition, SDSS is incomplete for low-mass QGs, while brief rejuvenation episodes make these systems spectroscopically detectable; thus, they are included in our sample. This selection bias would be muted in conventional SED fitting, because short ($\sim\!10\,\mathrm{Myr}$) star formation episodes outshine the older continuum and are smeared over multi-gigayear templates, which can produce spuriously high time-averaged sSFRs in massive systems.

In Figure~\ref{fig:color_evolution}, RGs lie blueward of the red sequence but do not linger in the green valley. Rejuvenation changes color rapidly at the onset of star formation, preventing a sizable steady-state population from forming in the valley. This can help explain the tension with \citet{Salim2014}, who searched for transition objects using NUV--$r$ color. NUV--$r$ is poorly suited to finding systems moving from the red sequence to the blue cloud because NUV is most sensitive to the most massive, shortest-lived stars. Under a rapid reignition of star formation, our \textsc{FSPS} models indicate that the passage through the green valley is $\lesssim\!1\,\mathrm{Myr}$.

We quantify the ratio of rejuvenating to quenched galaxies, $N_{\mathrm{RG}}/N_{\mathrm{QG}}$, as a function of group- or cluster-centric radius $r/r_{180}$ (Figure~\ref{fig:env_fraction}).
$N_{\mathrm{RG}}/N_{\mathrm{QG}}$ rises from $\sim\!2\%$ in rich-cluster cores to $\sim\!13\%$ at $r\!\sim\!r_{180}$ in group-mass halos and peaks in the field at $\sim\!16\%$. This trend suggests that low-density regions can supply or facilitate the retention of the fresh gas needed to rejuvenate quenched galaxies. By contrast, $N_{\mathrm{RG}}/N_{\mathrm{SF}}$ remains roughly constant at $\sim\!10\%$ in dense environments but drops to $\sim\!4\%$ in isolation, which indicates that inside halos, RGs make up a fixed share of the SFG population. In the field, they are diluted by the larger pool of continuously star-forming systems.

This pattern agrees with \citet{Levis2025}, who find that satellites rejuvenate primarily before crossing $r_{200}$ in \textsc{IllustrisTNG}. A simple two-stage sequence is consistent with both results. Pre-processing in smaller halos first quenches a galaxy. On approach to a group or cluster, minor mergers, large-scale structure, tidal torques, or ram-pressure stripping can deliver or destabilize gas, potentially reigniting star formation before the dense intracluster medium truncates the cycle.

\subsection{Gas Accretion as a Primary Rejuvenation Mechanism}\label{sec:metallicity}

\begin{figure}
  \centering
  \includegraphics[width=\columnwidth]{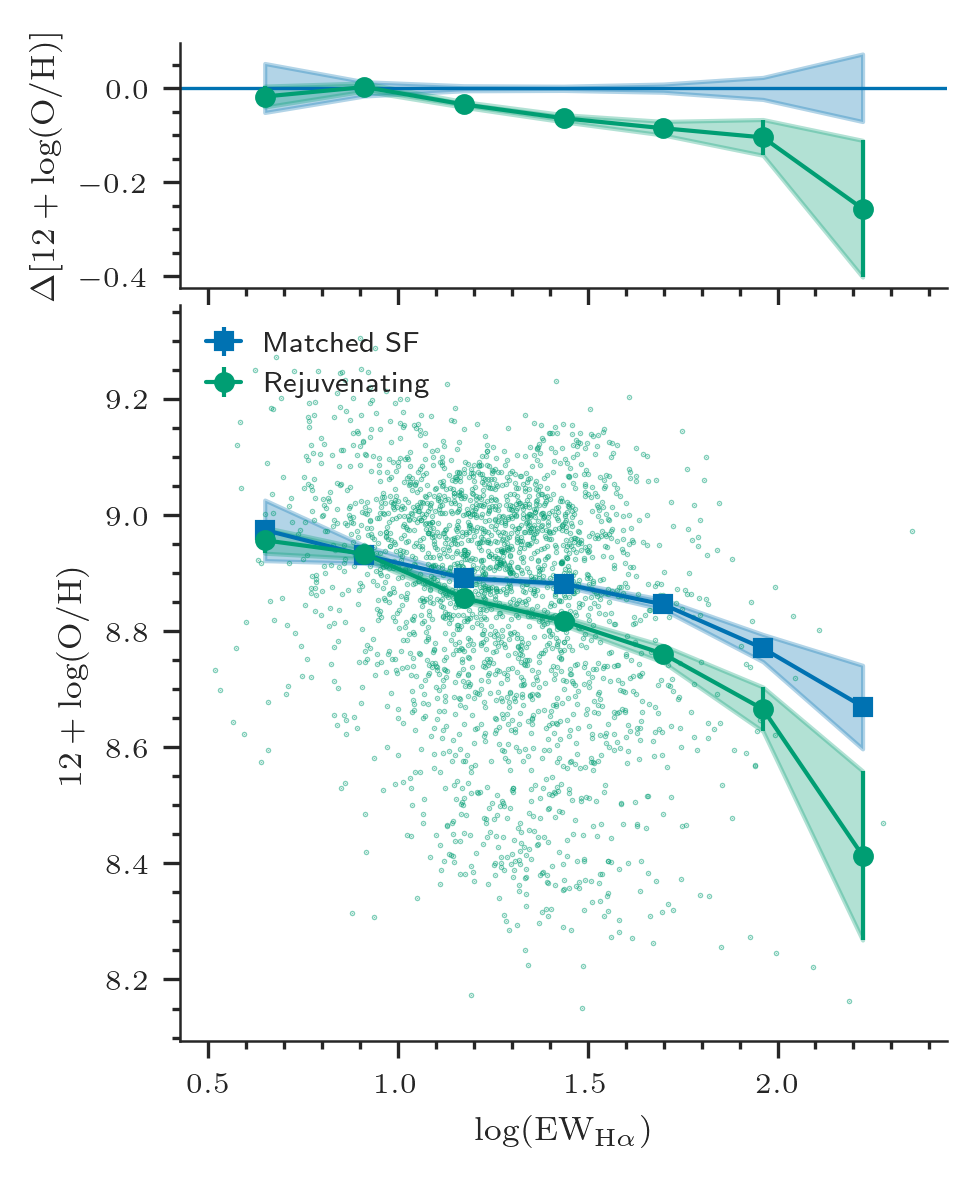}
  \caption{Lower: gas-phase metallicity vs. $\mathrm{EW}_{\Ha}$ for rejuvenating galaxies. Mean metallicities in bins of $\mathrm{EW}_{\Ha}$ are shown for the rejuvenating sample (green) and a mass-matched star-forming control (blue). Upper: metallicities of RGs relative to the mass-matched star-forming mean in each $\mathrm{EW}_{\Ha}$ bin.}
  \label{fig:metallicity}
\end{figure}

Gas-phase metallicity serves as an independent tracer of the fuel powering rejuvenation. To search for signatures of gas accretion, we compare \textsc{MPA--JHU} gas-phase metallicities between the SFG and RG populations, plotted against $\mathrm{EW}_{\Ha}$ in Figure~\ref{fig:metallicity}. We omit the quenching and quenched samples, as reliable metallicities require high-S/N lines. We mass-match the SFG control to the RG sample by dividing the SFG sample into 10 equal-width stellar-mass bins and randomly sampling SFGs with replacement until the RG counts are matched in each bin. We repeat this 100 times to mitigate random selection bias, then compute means for each sample across eight equal-width $\mathrm{EW}_{\Ha}$ bins. The rejuvenating population exhibits lower mean metallicities across nearly all bins, with the difference increasing toward higher $\mathrm{EW}_{\Ha}$, as highlighted by the residuals in the upper panel.

At fixed stellar mass and $\mathrm{EW}_{\Ha}$, rejuvenating galaxies are on average $\sim\!0.15\,\mathrm{dex}$ more metal poor than SFGs, increasing to $\sim\!0.25\,\mathrm{dex}$ for the highest-sSFR RGs. Such dilution is a classic signature of inflow of low-$Z$ gas or minor mergers with low-metallicity dwarfs—processes expected to peak in group outskirts, where satellite encounters are common and the intragroup medium has not yet been fully shock-heated \citep[e.g.,][]{Kaviraj2009,Keres2005}. This aligns with the increase of $N_{\mathrm{RG}}/N_{\mathrm{QG}}$ with $r/r_{180}$ (Figure~\ref{fig:env_fraction}), suggesting that the same mechanism that lowers metallicity can also supply a fresh reservoir of star-forming gas. Metal-poor fuel radiates less efficiently, reducing cooling rates and increasing depletion times, plausibly explaining why most RGs lie below the star-forming main sequence.

\section{Summary and Future Work}
\label{sec:summary}

We developed and applied a scalable technique to identify rejuvenating galaxies in the local Universe by combining GALEX UV photometry with SDSS \Ha\ spectroscopy. By (i) applying consistent dust corrections (IRX for NUV; Balmer decrement for \Ha) and (ii) predicting and subtracting the O-star--associated (ionizing) NUV from $L_{\mathrm{H}\alpha}$ using an \textsc{FSPS}-calibrated conversion, we cleanly separate the star formation timescales traced by NUV and \Ha. The result is a straightforward diagnostic that can be applied to large samples of photometric and spectroscopic data. Here we summarize our main results:

\begin{enumerate}[leftmargin=*]

    \item \textbf{Method (single-conversion + EPR flag).} Using a single equilibrium conversion (Equation~\ref{eq:eqconversion}), we predict and subtract the O-star--associated NUV from the dust-corrected NUV to isolate the $10$--$100\,\mathrm{Myr}$ component. We also define an early-phase rejuvenation (EPR) binary flag, set when $L_{\mathrm{NUV,non}}^{\mathrm{eq}} \le 0$, to tag the youngest, O-star--dominated galaxies within the rejuvenating class. 
    
    \item \textbf{Incidence.} We identify $\sim\!10^{4}$ rejuvenating galaxies ($\sim\!4.5\%$) in an SDSS parent sample of $223{,}703$ galaxies at $z<0.1$, showing that rejuvenation is uncommon but non-negligible in the local Universe.
    
    \item \textbf{Color evolution.} Rejuvenation quickly drives galaxies into the blue cloud, with only $14.6\%$ of RGs lying in the green valley at present (Figure~\ref{fig:color_evolution}).
    
    \item \textbf{Environment.} RGs are more common in groups than in the field or rich clusters, with fractions that increase toward group outskirts, suggesting that moderate densities facilitate gas replenishment.
    
    \item \textbf{Metallicity.} At fixed mass and $\mathrm{EW}_{\Ha}$, RGs have systematically lower gas-phase metallicities—on average $\sim\!0.15\,\mathrm{dex}$ below the star-forming control (rising to $\sim\!0.25\,\mathrm{dex}$ at the highest sSFR)—consistent with accretion of low-$Z$ gas.
    
\end{enumerate}

Together, these findings favor a rejuvenation channel linked to late-time gas inflows rather than purely stochastic variability in star formation. Our method can be extended to integral field unit surveys to map the spatial distribution of rejuvenation and its connection to kinematics and metallicity gradients, e.g., Mapping Nearby Galaxies at APO (MaNGA) \citep{bundy2015} and Sydney–AAO Multi-object Integral-field Spectrograph (SAMI) \citep{bryant2015}. Applications to higher-redshift samples will test the prevalence and drivers of rejuvenation beyond $z\!\sim\!0$ and help connect rejuvenation to gas supply, feedback, and the quenching cycle in galaxies.

\section*{Acknowledgments}
L.C.P. thanks the Natural Sciences and Engineering Research Council of Canada (NSERC) for funding.

Funding for the SDSS and SDSS-II has been provided by the Alfred P. Sloan Foundation, the Participating Institutions, the National Science Foundation, the U.S. Department of Energy, the National Aeronautics and Space Administration, the Japanese Monbukagakusho, the Max Planck Society, and the Higher Education Funding Council for England. The SDSS Web site is \url{http://www.sdss.org/}.

The SDSS is managed by the Astrophysical Research Consortium for the Participating Institutions. The Participating Institutions are the American Museum of Natural History, Astrophysical Institute Potsdam, University of Basel, University of Cambridge, Case Western Reserve University, University of Chicago, Drexel University, Fermilab, the Institute for Advanced Study, the Japan Participation Group, Johns Hopkins University, the Joint Institute for Nuclear Astrophysics, the Kavli Institute for Particle Astrophysics and Cosmology, the Korean Scientist Group, the Chinese Academy of Sciences (LAMOST), Los Alamos National Laboratory, the Max-Planck-Institute for Astronomy (MPIA), the Max-Planck-Institute for Astrophysics (MPA), New Mexico State University, Ohio State University, University of Pittsburgh, University of Portsmouth, Princeton University, the United States Naval Observatory, and the University of Washington.

This research is based on observations made with the Galaxy Evolution Explorer, obtained from the MAST data archive at the Space Telescope Science Institute, which is operated by the Association of Universities for Research in Astronomy, Inc., under NASA contract NAS 5–26555.

This publication makes use of data products from the Wide-field Infrared Survey Explorer, which is a joint project of the University of California, Los Angeles, and the Jet Propulsion Laboratory/California Institute of Technology, funded by the National Aeronautics and Space Administration.

\software{
\texttt{astropy} \citep{Astropy2013,Astropy2018,Astropy2022},
\texttt{matplotlib} \citep{Hunter2007},
\texttt{numpy} \citep{Harris2020},
\texttt{pandas} \citep{McKinney2010,Reback2020},
\texttt{scipy} \citep{Virtanen2020},
\texttt{seaborn} \citep{Waskom2021},
\texttt{scikit-learn} \citep{pedregosa_2011},
\texttt{FSPS} \citep{Conroy2009,Conroy2010,Byler2017},
\texttt{MIST} \citep{Choi2016,Dotter2016}}

\appendix

We provide a machine-readable table listing all rejuvenating galaxies; a short excerpt is shown.

\begin{deluxetable*}{lccccc}[h!]
\tabletypesize{\footnotesize}
\tablecaption{Rejuvenating Sample with EPR Flags\label{tab:mrt}}
\tablehead{
\colhead{SDSS \texttt{objID}} &
\colhead{R.A. ($^\circ$)} &
\colhead{Dec. ($^\circ$)} &
\colhead{$z \ ^{a}$} &
\colhead{$\log(M_\star/M_\odot) \ ^{b}$} &
\colhead{EPR flag$ \ ^{c}$}
}
\decimals
\startdata
1237658491730657444 & 152.704480 & 7.725634 & 0.097300 & 11.316000 & 0 \\
1237664338785927240 & 165.473370 & 36.327247 & 0.081300 & 11.315000 & 0 \\
1237652599027204457 & 318.689700 & -7.523477 & 0.065000 & 11.309000 & 1 \\
1237655469673546050 & 230.486090 & 2.344231 & 0.085600 & 11.306000 & 1 \\
1237663783658127559 & 354.591350 & -0.252282 & 0.090600 & 11.298000 & 0 \\
\multicolumn{6}{c}{\dots} \\
\enddata
\tablecomments{\\
$^{a}$ Spectroscopic redshift from the SDSS DR7 \citep{Abazajian2009}. \\
$^{b}$ SED-fitted stellar mass from the GSWLC-A2 \citep{Salim2016}. \\
$^{c}$ Early-phase rejuvenation flag; see Section~\ref{sec:method-main}. \\
(The full table is available in its entirety in machine-readable form in the online article.)
}
\end{deluxetable*}

\bibliography{references}
\bibliographystyle{aasjournal}

\end{document}